\documentclass[reprint,pra,superscriptaddress,twocolumn,floats,10pt,floatfix,nobibnotes]{revtex4-1}

\usepackage{amsmath}
\usepackage{amsfonts}
\usepackage{amssymb}
\usepackage{graphicx}
\usepackage{color}
\usepackage[colorlinks=true,citecolor=blue,linkcolor=blue,urlcolor=black]{hyperref}
\usepackage{times}
\usepackage{amstext}
\usepackage{latexsym}
\usepackage{bbm}

\newcommand{\bra}[1]{\langle #1|}
\newcommand{\ket}[1]{|#1\rangle}

\begin{document}

\title{Heralded entangling quantum gate via cavity-assisted photon scattering}

\author{Halyne S. Borges}
\email{halyneborges@gmail.com}
\author{Daniel Z. Rossatto}
\email{zini@df.ufscar.br}
\affiliation{Departamento de F\'{i}sica, Universidade Federal de S\~{a}o Carlos, 13565-905, S\~{a}o Carlos, S\~{a}o Paulo, Brazil}
\author{Fabr\'{i}cio S. Luiz}
\email{fabriciosouzaluiz@gmail.com} 
\affiliation{Instituto de F\'{i}sica de S\~{a}o Carlos, Universidade de S\~{a}o Paulo, 13560-970, S\~{a}o Carlos, S\~{a}o Paulo, Brazil}
\author{Celso J. Villas-Boas}
\email{celsovb@df.ufscar.br }
\affiliation{Departamento de F\'{i}sica, Universidade Federal de S\~{a}o Carlos, 13565-905, S\~{a}o Carlos, S\~{a}o Paulo, Brazil}


\begin{abstract}

We theoretically investigate the generation of heralded entanglement between two identical atoms via cavity-assisted photon scattering in two different configurations, namely either both atoms confined in the same cavity or trapped into locally separated ones. Our protocols are given by a very simple and elegant single-step process, whose key mechanism is a controlled-phase-flip gate implemented by impinging a single photon on single-sided cavities. In particular, when the atoms are localized in remote cavities, we introduce a single-step \textit{parallel quantum circuit} instead of the serial process extensively adopted in the literature. We also show that such parallel circuit can be straightforwardly applied to entangle two macroscopic clouds of atoms. Both protocols proposed here predict a high entanglement degree with a success probability close to the unity for the state-of-the-art parameters. Among other applications, our proposal and its extension to multiple atom-cavity systems step toward a suitable route for quantum networking, in particular for quantum state transfer, quantum teleportation and nonlocal quantum memory.
\end{abstract}

\maketitle

\section{Introduction}
\label{sec:1}

Quantum mechanics is driving forward a technological revolution in the $21$st century, overcoming the miniaturization barrier and the performance of devices that can be achieved within a classical framework \cite{Milburn2003}, i.e., achieving the so-called quantum supremacy \cite{Preskill2012}. At the heart of this revolution are the quantum correlations, in particular the quantum entanglement \cite{Horodecki2009}. Besides being an unique element of quantum mechanics that has no a classical counterpart \cite{Brunner2014}, entanglement is a cornerstone of several quantum devices and protocols. For instance, it is used to implement quantum logical gates \cite{Nielsen2000, Rauss2001}, to perform more efficient computation algorithms \cite{Montanaro2016,Terhal2015}, and to share secure information \cite{Gisin2007,Scarani2009} in quantum computers and networks \cite{Kimble2008,Ladd2010,Meter2014}. Therefore, entanglement plays an extremely important role in both fundamental and applied physics, so that quantum devices that efficiently generates it are highly desirable.

Nonetheless, only efficient generation of entanglement is not enough. These devices must also be scalable and robust against decoherence \cite{Milburn2003}. Hybrid systems composed of photons and atoms trapped into cavities (resonators) in the strong-coupling regime are excellent candidates to meet these requirements \cite{Auffeves2013}. In this scenario, atomic systems can be better isolated from the effects of the environment and connected at long distances through optical photons, thus forming elementary quantum networks that use photons to distribute entanglement \cite{Ritter2012}. This so-called cavity-based quantum network is a very active research field and is auspicious for quantum networks on larger scales, framework in which theoretical and experimental progress have been made in quantum computing and communication \cite{Reiserer2015}.

Inside this research field, the quantum mechanism that is the building block of our proposal consists of a controlled-phase-flip (CPF) gate performed via cavity-assisted photon scattering (CAPS), i.e., the gate is performed by impinging a single-photon pulse on a single-sided cavity, which is coupled to a single \cite{Duan2004} (or $N$ \cite{Duan2005}) three-level atom. 
Since a CPF gate together with simple single-qubit operations performs universal quantum computation \cite{Preskill}, it has attracted much interest in the last decades \cite{Xiao2004,Cho2005,Lin2006,Lin2006-2,Deng2007,Deng2007-2,Wei2007,xialong2008,Hu2008,Lee2008,An2009,Song2009,Chen2011,Borges16}. Furthermore, the recent experimental achievements in CAPS-based CPF gates \cite{Reiserer2013,Reiserer2014,Kalb2015,Hacker2016} have put this subject in the scientific spotlight again.

In particular, some of these referred works theoretically propose schemes to entangle atoms localized in long-distance cavities via CAPS \cite{Cho2005,Lin2006,Lin2006-2,Deng2007,Deng2007-2,Hu2008,Lee2008,An2009,Song2009,Chen2011}. In contrast to other approaches for entangling distant atoms that require both the interference and the simultaneous detection of two photons emitted from the two respective atoms \cite{Barret2005,Moer2007,Hoff2012,Bernien2013,Casabone2013,Delteil2016}, or require that one atom absorbs a single photon emitted by the other atom \cite{Ritter2012}, the CAPS-based protocols have the advantages of requiring only a single photon and not requiring an energy exchange between the parties. However, even though some of the aforementioned CAPS-based entangling gates are given by a single-step process, the single-photon pulse impinges on the cavities in sequence ({\it serial quantum circuit}) until being detected at the end. Moreover, single-qubit operations are performed in the atoms and/or the pulse during the process.

In this work, we theoretically investigate two protocols of CAPS-based heralded-entanglement generation between two atoms. We consider here that our flying qubit (single-photon pulse) is encoded by the vacuum and single-photon states instead of the polarizations of the pulse widely adopted in the related literature. This choice has an advantage of performing fewer (or even no) single-qubit operations on the flying qubit during the process. In the first protocol, we consider both atoms placed in the same cavity and analyze the atomic entanglement that is generated by impinging a single photon on the cavity. Although this configuration was already introduced in Ref.~\cite{Duan2005}, a specific analysis regarding the entanglement generation between the atoms is missing. Here we perform such analysis, whereby we provide semianalytical results for the entanglement degree acquired by the atoms and for the success probability of measuring the outgoing photon, which heralds the entangling gate. We show that both the entanglement and the success probability are very close to the unity for the current technology. In addition, our study allows us to make a comparison between the efficiency of the entangling gate investigated here and another similar one recently carried out, which carves the atomic state by measuring photon pulses reflected by the cavity \cite{Welte2017}.

Subsequently, we propose a protocol to entangle two atoms localized in remote cavities by using a kind of single-step \textit{parallel quantum circuit} instead of the serial process extensively adopted in the literature. In this case, a single-photon pulse crosses a $50$:$50$ beam splitter (BS), such that it virtually impinges on both cavities at same time, with the outgoing pulse being detected after passing again through the BS. Likewise the first case, this protocol also provides an entanglement degree and a total success probability very close to the unity for the state-of-the-art parameters. Finally, we also show that this parallel circuit can be straightforwardly applied to entangle two distant macroscopic clouds of atoms through the same simple single-step process.

Although we specifically use the optical domain and atomic systems in this work, it is worth stressing that the process of our entangling gates could also be adapted and further developed in solid-state-based systems that employ similar techniques and concepts, such as superconducting circuits \cite{Devoret2013} and quantum dots \cite{Petta2013,Lodahl2015}.


\section{Physical system and model}
\label{sec:2}

We investigate the entanglement generation between a pair of identical atoms either confined into the same single-mode cavity or trapped into long-distance cavities. Each atom is described as a three-level system in a $\Lambda$-level configuration, in which the excited state $|3\rangle$ is resonantly coupled to the ground one $|1\rangle$ through the cavity mode, while the other ground state $|2\rangle$ remains decoupled, as shown in Fig.~\ref{fig:1}.  Here, we consider that the cavity has only one partially transmitting mirror (single-sided cavity) that couples the intracavity mode to the continuum of free-space modes, which can be considered as a bosonic reservoir. Hence, an incoming field could only enter and then exit from the cavity by one side. In both configurations, the initial state of the system is given by each atom in a balanced superposition of its ground states and the cavity (or cavities) in the vacuum state. Furthermore, the external multimode field of the bosonic reservoir initially has its excited modes centered on the cavity resonance frequency, i.e., we consider an incoming single-photon pulse (input field) as a quasimonochromatic wave packet whose spectral spread is much smaller than the carrier frequency \cite{Loudon}, which is exactly the cavity resonance frequency in our case.

In our proposal, the entanglement generation protocol has the implementation of a CPF gate as a key ingredient, with which an induced phase can be imprinted in the output field depending on the atomic state \cite{Duan2004, Duan2005, Borges16}. As said before, only the atomic transition $|1\rangle \leftrightarrow |3\rangle$ is coupled through the cavity mode. Therefore, when the cavity-atom system is in the strong-coupling regime, the normal-mode splitting ensures that, if the atom is in $|1\rangle$, the incident photon is immediately reflected, having its phase changed by an amount of $\pi$. On the other hand, if the atom is in $|2\rangle$, it becomes transparent to the cavity field, such that the input field enters the cavity and then is transmitted without any change in its phase \cite{phase}. For $N$ atoms inside the cavity, the situation is similar with the output photon not acquiring a phase shift only when all atoms are in $|2\rangle$, but acquiring a phase of $\pi$ otherwise. For more details about the implementation of this CPF gate, see Refs.~\cite{Duan2004, Duan2005, Borges16}.   

In the following we introduce our protocols of entanglement generation and individually analyze each configuration, but without restricting to the strong-coupling regime for the cavity-atom system.

\subsection{Two atoms inside the same cavity} 
\label{sec:two}

\begin{figure}[t]
\centering
\includegraphics[width=0.45\textwidth]{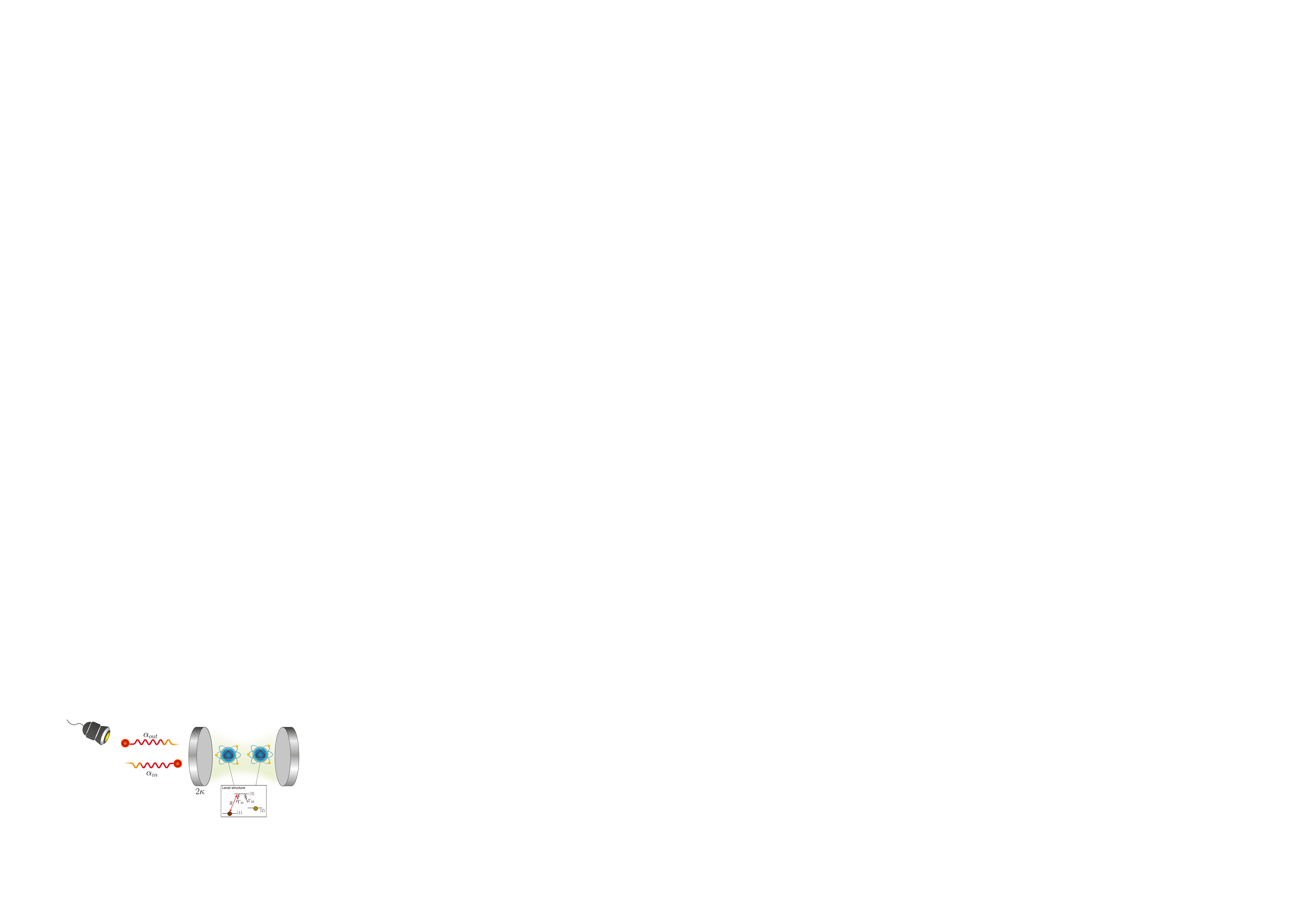}
\caption{Pictorial representation of the quantum system for two noninteracting atoms confined into the same single-sided cavity, which has a decay rate of $2\kappa$. Each atom is described as a three-level system in a $\Lambda$-configuration, whose atomic transition $|1\rangle \leftrightarrow |3\rangle$ is resonantly coupled to the intracavity mode with  coupling strength $g$. The rates $\Gamma_{31}$ and $\Gamma_{32}$ describe the spontaneous emission from the atomic excited state to the ground ones. Here, an input field ($\alpha_{\text{in}}$) impinges on the partially transmitting mirror of the cavity, such that a detector can register a photon count in the output field ($\alpha_{\text{out}}$) when there is no photon loss via atomic spontaneous emission.}
\label{fig:1}
\end{figure}

In this setup, the pair of noninteracting three-level atoms are confined into the same single-sided cavity, with only the atomic transition $|1\rangle \leftrightarrow |3\rangle$ being coupled to the intracavity mode with coupling strength $g$. A pictorial representation of the system assisted by the input ($\alpha_{\text{in}}$) and output ($\alpha_{\text{out}}$) fields is illustrated in Fig.~\ref{fig:1}, where $2\kappa$ stands for the cavity decay rate that determines the out-coupling of the external modes and the intracavity one through the partially transmitting mirror.

Considering the case in which are valid the weak-coupling approximation between the cavity and its reservoir (Born approximation), the rotating-wave and the Markov approximations, i.e., the so-called white-noise limit \cite{Zoller2004}, the Hamiltonian that describes the entire system in an interaction picture rotating at the cavity resonance frequency is ($\hbar =1$) \cite{Zoller2004,Walls2007,Lukin2000}
\begin{align} \label{eq:1}
H &= \int_{-\infty}^{\infty} \!\!\!\! d\omega \, \omega b^{\dagger}(\omega)b(\omega)+ i\frac{\sqrt{2\kappa}}{\sqrt{2\pi}}\int_{-\infty}^{\infty} \!\!\!\! d\omega \,[a^{\dagger}b(\omega) - ab^{\dagger}(\omega)] \nonumber \\
&+\sum_{j=A,B}g(a\sigma_{31}^{j} + a^{\dagger}\sigma_{13}^{j}),
\end{align} 
in which $b(\omega)$ is the frequency-dependent annihilation operator of the bosonic reservoir $([b(\omega),b^{\dagger}(\omega ^{\prime})]=\delta(\omega - \omega^{\prime}))$, $a$ is the annihilation operator of the intracavity mode, and $\sigma_{k\ell}^{j} = | k \rangle \langle \ell |$ is a ladder operator of the $j$-th atom.

We are interested in an initial state with only one excitation in the input field of the following kind
\begin{equation} \label{eq:2}
|\psi(t_0)\rangle = \underbrace{[(\ket{1}+\ket{2})/\sqrt{2}]_{AB}^{\otimes 2}}_{\text{atoms $A$ and $B$}} \otimes \underbrace{\ket{0}_{c}}_{\text{cavity}} \otimes \underbrace{\ket{\mathbf{1}_{\xi}}}_{\text{reservoir}},
\end{equation}
in which $|\mathbf{1}_\xi \rangle = \int_{-\infty}^{\infty} d\omega \xi_{\text{in}} (\omega) b^{\dagger}(\omega) |\mathbf{0} \rangle$ describes the input field in a continuous-mode single-photon state (in the interaction picture concerned \cite{rotcarr}), which can be interpreted as a single photon coherently superposed over many spectral modes with weighting given by the spectral density function $\xi_{\text{in}} (\omega)$ \cite{Loudon,Milburn2008}. Here, $|\mathbf{0} \rangle$ is the continuous-mode vacuum state ($b(\omega)|\mathbf{0} \rangle = 0$) and from the normalization condition we have $\int_{-\infty}^{\infty} d\omega |\xi_{\text{in}} (\omega)|^2 = 1$. The Fourier transform of $\xi_{\text{in}}(\omega)$ provides the square-normalized temporal shape $\alpha_{\text{in}}(t)$ of the incoming single-photon pulse ($\int_{-\infty}^{\infty} dt |\alpha_{\text{in}} (t)|^2 = 1$). Hereafter, we will omit the tensor product, $\otimes$, for the sake of simplicity.

In the ideal case ($g \to \infty$), the CPF gate provides $\ket{k\ell}_{AB} \ket{0}_{c} \ket{\mathbf{1}}_{\text{in}} \to -(-1)^{(k-1)(\ell-1)} \ket{k\ell}_{AB}  \ket{0}_{c} \ket{\mathbf{1}}_{\text{out}}$ for $k,\ell \in \{ 1,2\}$ \cite{Duan2005,Borges16}. Thus, for the initial state given by  Eq.~\eqref{eq:2} we obtain after applying the CPF gate
\begin{equation} \label{psi_ideal}
|\psi_{g\to\infty}\rangle =\frac{1}{\sqrt{2}}(\ket{2-}-\ket{1+})_{AB} \ket{0}_{c} \ket{\mathbf{1}}_{\text{out}},
\end{equation}
with $\ket{\pm} = (\ket{2} \pm \ket{1})/\sqrt{2}$. Therefore, for sufficiently strong $g$, the atoms which are initially factorized, become maximally entangled just by implementing a very simple single-step process, i.e., just by impinging a single-photon pulse on the cavity. It is worth mentioning that any balanced superposition between the atomic ground states can be used as the atomic initial state. In the following we investigate in more detail this entanglement generation, extending the results discussed in Ref.~\cite{Duan2005}.

Given the initial state [Eq.~\eqref{eq:2}] and the Hamiltonian [Eq.~\eqref{eq:1}] in question, the general evolved state can be written in the form:
\begin{align} \label{psi_t}
\ket{\psi(t)} &= \sum_{k=1}^{2} \left( c_{k3,0}(t) \ket{k3}_{AB} + c_{3k,0}(t) \ket{3k}_{AB}  \right) \ket{0}_c  \ket{\mathbf{0}}  \nonumber \\
 & + \sum_{k,\ell=1}^{2}  c_{k\ell,1}(t)\ket{k\ell}_{AB} \ket{1}_{c}  \ket{\mathbf{0}}  \nonumber \\
 & + \sum_{k,\ell=1}^{2}\int_{-\infty}^{\infty} \!\!\!\! d\omega \, \xi_{k\ell}(\omega,t)\ket{k\ell}_{AB} \ket{0}_{c} \, b^{\dagger}(\omega)\ket{\mathbf{0}}.
\end{align}
Nevertheless, the atomic spontaneous emission is an inevitable incoherent process. In our case, this process would lead to either the transitions $\ket{3} \rightarrow \ket{1}$ or $\ket{3} \rightarrow \ket{2}$ of one of the atoms at the expenses of a photon loss to the free-space external modes orthogonal to those parallel to the cavity axis, which in turn are coupled to the intracavity mode. These transitions lead to a vacuum-state output field with the system deexcited, yielding a leakage error on the CPF gate since the final state is outside of the desired Hilbert space (states in Eq.~\eqref{psi_t}) \cite{Duan2004,Duan2005,Preskill}. However, the gate errors due to all sources of photon loss \cite{phloss} can always be indicated when a detector (see Fig.~\ref{fig:1}) does not register a photon count from the output field. This dominant noise generates a failure probability of the CPF gate, but it does not affect the gate fidelity if the operation succeeds (if the photon is not lost) \cite{Duan2004,Duan2005}. Thus, a photon count is not mandatory to perform the CPF gate, but it heralds its success without introducing any perturbation to the gate, i.e., a photon detection just assures us if the gate has succeeded or not. For this class of probabilistic signaled errors, efficient quantum computation is possible even for an arbitrarily small gate success probability \cite{Duan2005-2, Duan2008}, postselecting those cases in which a photon is detected. Fortunately, the success probability of our entangling gate is very close to the unity for the current technology as we will show.

Focusing only on the cases in which a photon is never lost, the atomic spontaneous emission can be phenomenologically taken into account by introducing non-hermitian damping terms into the interaction Hamiltonian \cite{Duan2004,Duan2005}
\begin{equation} \label{heff}
H_{\text{eff}} \rightarrow H - i\Gamma_{3}\sum_{j}\sigma_{33}^{j},
\end{equation}
in which $\Gamma_{3} = \Gamma_{31} + \Gamma_{32}$, with $\Gamma_{31}$ and $\Gamma_{32}$ standing for the decay rates from the excited state $|3\rangle$ to the ground states $|1\rangle$ and $|2\rangle$, respectively. In this way, the dynamics evolves only inside the desired Hilbert space, such that the general state given by Eq.~\eqref{psi_t} still holds. On the other hand, due to the non-hermitian terms in $H_{\text{eff}}$, the Schr\"{o}dinger equation will now provide an unnormalized $\ket{\psi(t)}$ whose squared norm $(|\langle \psi(t) \ket{\psi(t)}|^{2} )$ exactly gives the probability of not losing a photon by atomic spontaneous emission in the time interval between $t$ and $t + dt$.

When the outgoing field is far enough from the cavity ($t \to \infty$), i.e., when the pulse and the atoms-cavity system no longer interact, the normalized atomic steady state predicted by the Schr\"{o}dinger equation, $ i \partial_{t}\ket{\psi(t)} = H_{\text{eff}}\ket{\psi(t)}$ ($\hbar =1$), after the detector registering a photon count, is given by (see Appendix \ref{appA})
\begin{equation} \label{rhoatss}
\rho_{\text{at}}^{\text{ss}} = \frac{1}{P_s}\sum_{k,\ell=1}^{2} \sum_{p,q=1}^{2} \int_{-\infty}^{\infty} \!\!\!\! dt \, \alpha_{\text{out}}^{k\ell}(t) \, \alpha_{\text{out}}^{pq}(t)^{*} \ket{k\ell}\bra{pq},
\end{equation}
in which ${\alpha_{\text{out}}^{k\ell}}(t)$ is determined by the input-output relation, $\alpha_{\text{out}}^{k\ell}(t) = \sqrt{2\kappa} \, c_{k\ell,1}(t) - \alpha_{\text{in}}^{k\ell}(t)$, and 
\begin{equation}
P_s \equiv \sum_{k,\ell=1}^{2} \int_{-\infty}^{\infty} \!\!\!\! dt \, |\alpha_{\text{out}}^{k\ell}(t)|^{2},
\end{equation}
which is exactly the average number of photons in the output field after a long time, i.e., it gives the probability of registering a photon count (neglecting the detector inefficiency).

For evaluating and analyzing the degree of entanglement between the atoms (concurrence $\mathcal{E}(\rho)$ \cite{Wootters01}), we assume, without loss of generality, $\Gamma_{3} = \kappa$ in order to provide a compact semianalytical solution for $c_{k\ell,1}(t)$, which yields
\begin{align}\label{aout_analy}
{\alpha_{\text{out}}^{k\ell}(t)} = 2\kappa \! &\int_{-\infty}^{t} \!\!\!\! ds \,  \cos{[g \sqrt{\delta_{k,1}+\delta_{\ell,1}}(t-s)]}\nonumber \\
&\times \alpha_{\text{in}}^{k\ell}(s)e^{-\kappa(t-s)} - \alpha_{\text{in}}^{k\ell}(t).
\end{align}
This choice does not restrict our results, since one can show (numerically at least) that, after all, $\mathcal{E}(\rho_{\text{at}}^{\text{ss}})$ and $P_s$ depend only on the cooperativity parameter, $C\equiv g^{2}/2\kappa\Gamma_{3}$, regardless the combination of ($g,\kappa,\Gamma_{3}$) that provides the same $C$. Moreover, we consider an input pulse with a Gaussian temporal shape
\begin{equation}
\alpha_{\text{in}}(t)=\frac{1}{\sqrt{\eta\sqrt{\pi}}}e^{-\frac{1}{2}\frac{(t-t_{0})^{2}}{\eta^{2}}},
\label{shape}
\end{equation}
whose full width at half maximum that determines the pulse duration is $\tau_p=2\eta \sqrt{2\ln (2)}$, and whose maximum impinges on the cavity semitransparent mirror at $t_0$. In this way, the system reaches its steady state when $t \gtrsim t_0 + \tau_p$.

\begin{figure}[t]
\includegraphics[trim = 0mm 8mm 0mm 9mm, clip, width=0.48\textwidth]{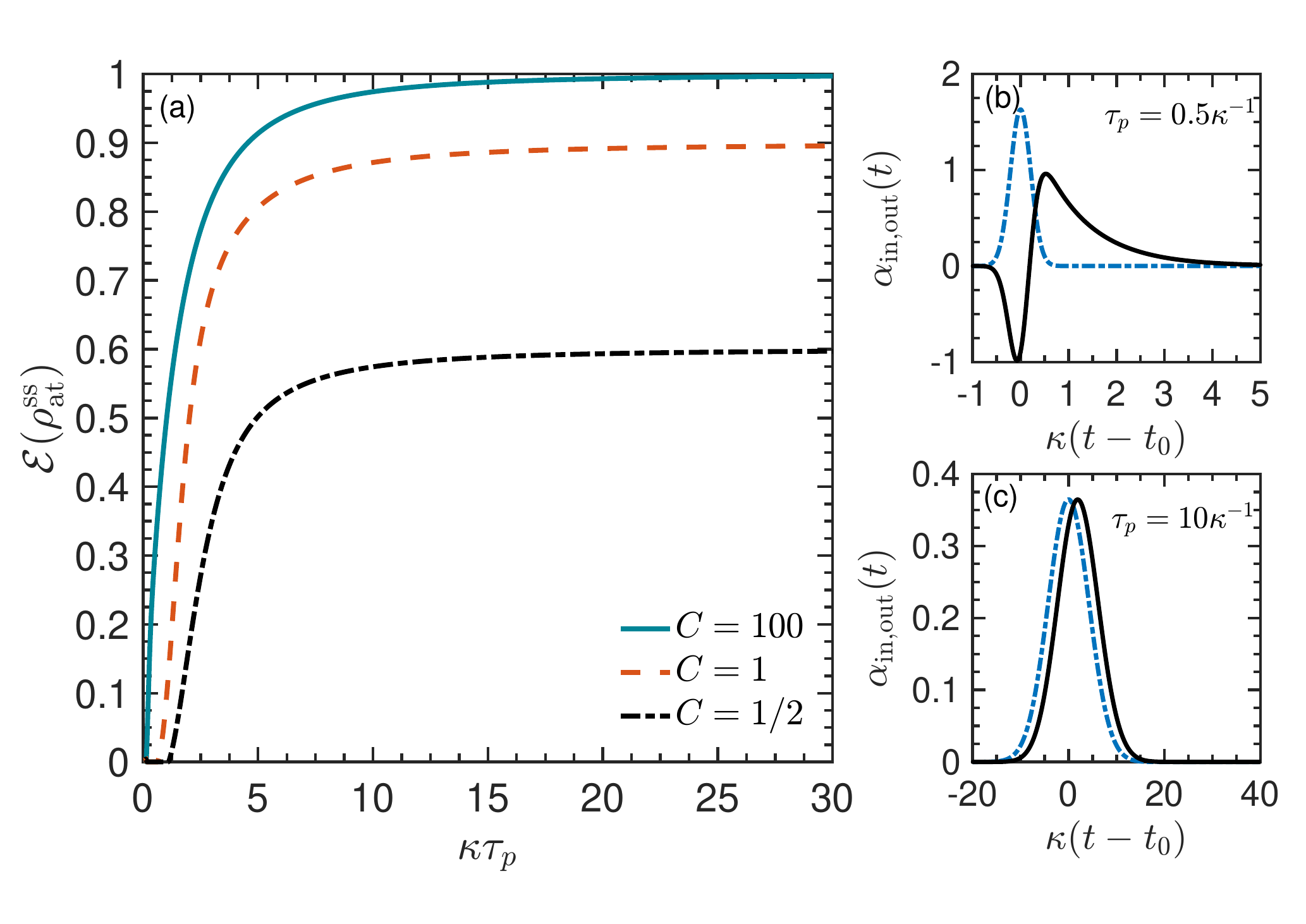}
\caption{(a) Concurrence ($\mathcal{E}$) of the atomic steady state ($\rho_{\text{at}}^{\text{ss}}$) as a function of pulse duration ($\tau_p$) considering different values of cooperativity ($C$) and using the initial state of Eq.~\eqref{eq:2}. Shape in the time domain of the input ($\alpha_{\text{in}}$ $-$ dashed line) and output ($\alpha_{\text{out}}$ $-$ solid line) fields for (b) short ($\tau_p = 0.5\kappa^{-1}$) and (c) long $(\tau_p = 10\kappa^{-1}$) input pulses, when both atoms are in the state $|2\rangle$ (empty-cavity-like case).}
\label{fig:2}
\end{figure}

Aside from the preparation time of the initial state as well as the propagation time of the input and output pulses, the required time to perform the CPF gate is given by the time interval over which the single-photon pulse interacts with the atoms-cavity system, which is dictated by the pulse duration $\tau_p$. Thus, the entanglement generation would be faster for shorter pulses. However, for a fixed $C$, the shorter $\tau_p$ the smaller the atomic entanglement degree ($\mathcal{E}$) generated in the steady state considering the initial state of Eq.~\eqref{eq:2} ($\alpha_{\text{in}}^{k\ell}(t)=\alpha_{\text{in}}(t)/2$), as shown in Fig.~\ref{fig:2}(a). This occurs because of a mismatch between the shapes of the input and output pulses for short $\tau_p$, which reduces the CPF gate fidelity \cite{Duan2004,Duan2005,Borges16} and, consequently, the generated entanglement. In this case, the deformation in the shape of the output pulse is due to the fact that a part of the input pulse is directly reflected ($\pi$ phase shift) by the system regardless the atomic state, while the other one is absorbed and then transmitted (no phase shift) depending on the initial atomic state and the atom-field coupling strength.

In  Figs.~\ref{fig:2}(b)$-$(c), we illustrate an example considering both atoms initially in the state $|2\rangle$ [$\alpha_{\text{in}}^{k\ell}(t) = \delta_{k,2}\delta_{\ell,2}\alpha_{\text{in}}(t)$], which is equivalent to the case of an empty cavity. For a long $\tau_p$, the spectral spread of the input pulse ($\tau_p^{-1}$) fits into the linewidth of the cavity ($2\kappa$), so that the whole input pulse enters and then exits from the cavity, yielding an output pulse that preserves the shape of the input pulse [Fig.~\ref{fig:2}(c)]. If the pulse is short ($\tau_p^{-1} \gtrsim 2\kappa$), part of the input pulse is out of resonance with the cavity (the spectral spread of the input pulse has a frequency interval that exceeds the cavity linewidth), so that this part is directly reflected by the cavity mirror while the other part is absorbed and then transmitted. Since it might has a delay between the reflected and transmitted parts in this case, and since they acquire opposite phase shifts, the shape of the output pulse gets deformed in comparison to the input-pulse shape  [Fig.~\ref{fig:2}(b)]. There exists an approximate lower bound for the pulse duration ($\tau_{\ket{2}}$) above which the desired pairing between the input- and output-pulse shapes occurs with high fidelity when both atoms are in $\ket{2}$. As we will briefly discuss in the next paragraph and can be seen in Fig.~\ref{fig:2}(a), the gate fidelity when at least one of the atoms are in $\ket{1}$ depends not only on the pulse duration, but also the atoms-cavity interaction and, consequently, $C$. Even so, we can assume $\tau_{\ket{2}} \simeq 10\kappa^{-1}$, since in Fig.~\ref{fig:2}(a) we note that $\mathcal{E}(\rho_{\text{at}}^{\text{ss}})$ approximately saturates for $\kappa\tau_p \gtrsim 10$ regardless the value of $C$.

When at least one of the atoms is in the state $\ket{1}$, the greater $C$ and the longer $\tau_p$, the higher the fidelity of the gate, as shown in Fig.~\ref{fig:2}(a). For this case, the desired pulse reflection happen when the input pulse is entirely out of resonance with the normal modes of the system (dressed states), i.e., when no portion of the spectral spread of the input pulse fits into the linewidth of the normal modes. This can be indeed achieved in the strong-coupling regime ($C \gg 1$), where there is a large normal-mode splitting, together with the use of long pulses ($\tau_{p} \gtrsim \tau_{\ket{2}}$).

\begin{figure}[t]
\includegraphics[trim = 11mm 8mm 14mm 12mm, clip, width=0.48\textwidth]{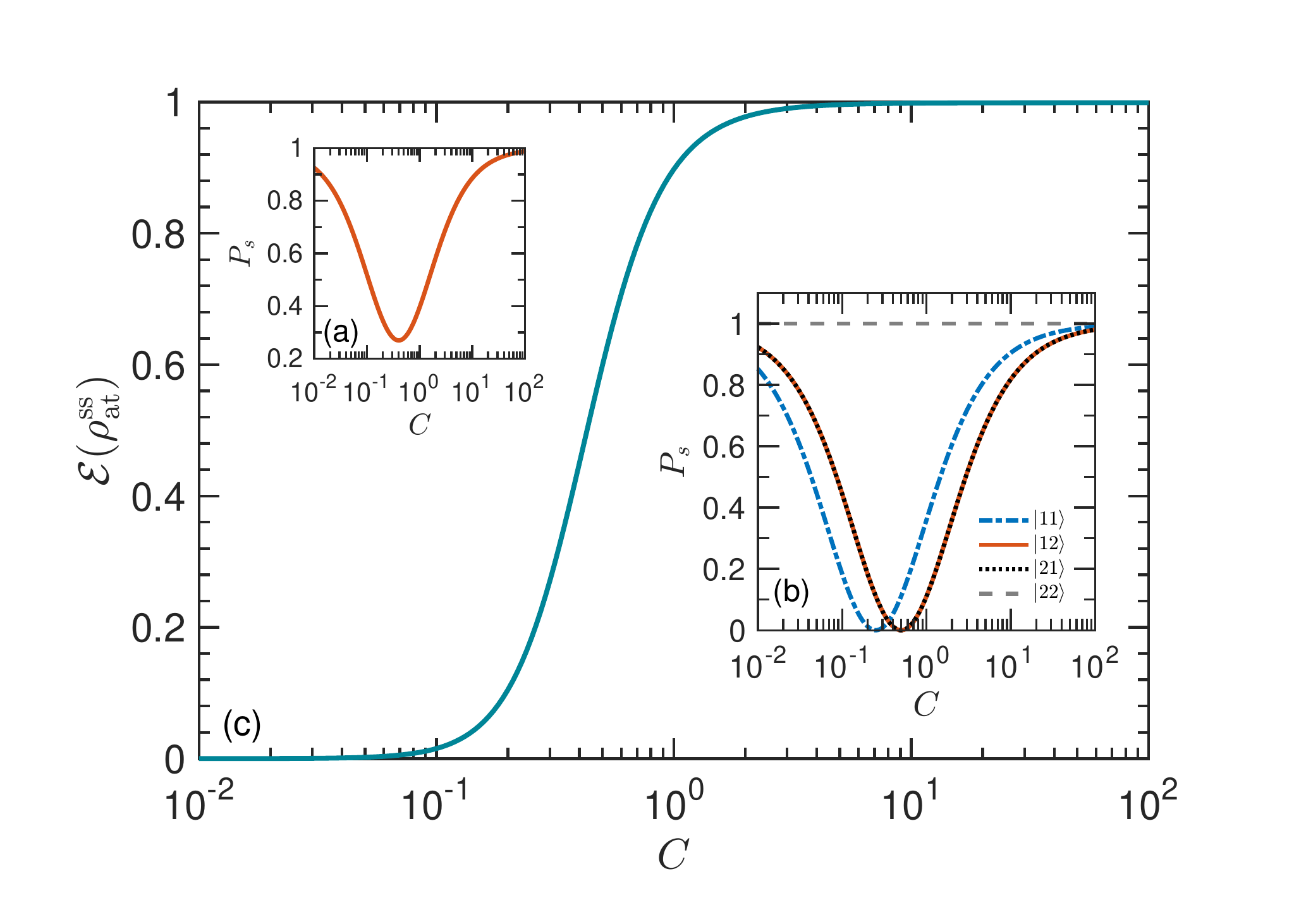}
\caption{(a) Average number of photons outside the cavity ($P_s$) and (c) the atomic concurrence as a function of the cooperativity in the steady regime, considering the atom initially in the state given by Eq.~\eqref{eq:2} and the $\tau_p=50\kappa^{-1}$. In (b) we show the contribution of each one of the separable atomic states in $P_s$ when they are initially prepared.}
\label{fig:4}
\end{figure}

Based on the above discussions together with the fact that we are interested to know on which conditions the degree of the entanglement can be optimized, hereafter we consider long pulses only. In Fig.~\ref{fig:4} we illustrate the average number of photons outside the cavity ($P_s$) and the concurrence as a function of the cooperativity in the steady regime. In Fig.~\ref{fig:4}(a) it is shown $P_s$ for the desired initial state given by Eq.~\eqref{eq:2}, while in Fig.~\ref{fig:4}(b) we consider each one of the four atomic state $|ij\rangle$ ($i,j \in \{1,2\}$) as the initial state, allowing us to analyse the individual contribution in the desired initial state [Eq.~\eqref{eq:2}]. 

Looking at Fig.~\ref{fig:4}(b) we can note that, when both atoms are in $|2\rangle$, the incoming pulse is always transmitted by the cavity regardless the value of $C$. In contrast, when at least one atom is in $|1\rangle$, the single photon that enters in the cavity has a probability to be lost by atomic spontaneous emission, resulting in a decrease of the probability of registering a photon count. For a specific value of $C$ the single photon is always lost ($P_s = 0$). When there is only one atom in $|1\rangle$ this happens for $C=1/2$ \cite{Borges16}, while for $C=1/4$  when both atoms are in $|1\rangle$. In general, $P_s=0$ when $C_M \equiv MC=1/2$, in which $M$ is the number of atoms in $\ket{1}$. For the desired initial state [Eq.~\eqref{eq:2}], since $P_s$ takes into account the contribution of the four states equally, it never cancels out, as we can notice in Fig.~\ref{fig:4}(a). Another interesting feature is that the atoms get entangled even within the region in which the CPF gate cannot be performed ($C_M \le 1/2$ \cite{Borges16}), such that it can unbalance the initially equal contribution of each separable atomic state without imprinting any phase shift on them, but being still able to entangle the atoms with certain degree. Therefore, we have that the degree of the entanglement is non-null even for these values of $C$ for which the CPF cannot work, as illustrated in Fig.~\ref{fig:4}(c).

We can also notice from Fig.~\ref{fig:4}(c) that the entanglement monotonically increases with the cooperativity, with the atoms becoming almost maximally entangled without requiring too strong coupling neither too high cooperativity, namely $\mathcal{E}(\rho_{\text{at}}^{\text{ss}}) \gtrsim 0.99$ for $C \gtrsim 3$, but both are beneficial for high fidelities and high efficiencies, since \{$\mathcal{E},P_s\} \to 1$ when $C \gg 1/2$. Here we consider a Gaussian shape for the input pulse, but it is worth stressing that its exact shape is actually not important if the shape changing is sufficiently slow compared to the cavity decay rate~\cite{Duan2004}. It is  also worth stressing that for real detectors the success probability of our entangling gate will be simply reduced by a factor due to detection efficiency.

Recently, Welte {\it et al.} \cite{Welte2017} experimentally demonstrate entanglement generation of two neutral atoms trapped inside an optical cavity. The authors essentially use the same setup of Fig.~\ref{fig:1}, but the entanglement is generated by carving the atomic state through the detection of a few weak photon pulses reflected from the cavity \cite{sorensen2003}. They achieved experimental parameters equivalent to $C \simeq 4.1$, which yields a success probability of $32\%$ ($50\%$ for their ideal scenario) for entangling the atom with high fidelity. In our case, such cooperativity yields $P_s \simeq 75\%$ for our ideal scenario. Besides, our entangling gate has another advantage since it requires the detection of only one photon pulse, avoiding the accumulation of errors due to dark counts.

\subsection{Two atoms in remote cavities}
\label{sec:twodis}
\begin{figure}[t]
\includegraphics[width=0.45\textwidth]{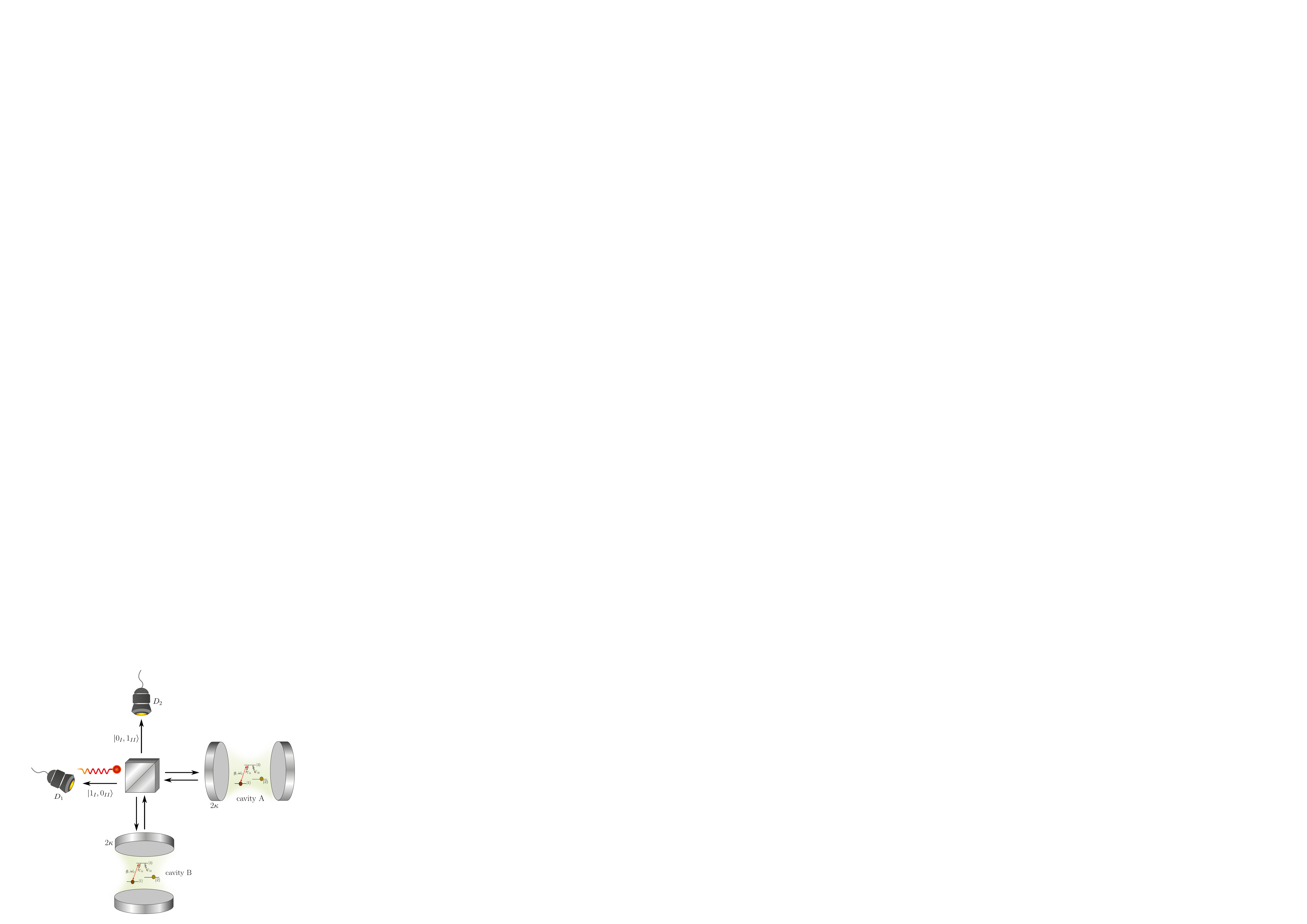}
\caption{Pictorial representation of the experimental setup when the atoms are trapped into locally separated cavities. A single photon crosses a $50$:$50$ BS, virtually impinges on both cavities at the same time, and then is detected by one of the detectors ($D_1$ or $D_2$) after passing through the BS again.}
\label{fig:5}
\end{figure}

Now let us consider the case in which the atoms are no longer inside the same cavity, but trapped into locally separated cavities, as depicted in Fig.~\ref{fig:5}. Again we assume the atoms initially in a balanced superposition ($\ket{+}$) with the cavities in the vacuum state. Our protocol to entangle the atoms firstly consists of sending a single-photon pulse to a 50:50 beam splitter (BS) through its channel I [see Fig.~\ref{fig:5}], so that we have $\ket{\mathbf{1}_\text{I},\mathbf{0}_\text{II}}\ket{\mathbf{0}_\text{III},\mathbf{0}_\text{IV}}$ initially. After crossing the BS, the photon has equal probability ($50\%$) to be transmitted through the channel III or reflected through the channel IV, namely \cite{Campos1989}
\begin{equation}
\text{BS} \ket{\mathbf{1}_\text{I},\mathbf{0}_\text{II}}\ket{\mathbf{0}_\text{III},\mathbf{0}_\text{IV}} = \ket{\mathbf{0}_\text{I},\mathbf{0}_\text{II}}\frac{(\ket{\mathbf{1}_\text{III},\mathbf{0}_\text{IV}}+i\ket{\mathbf{0}_\text{III},\mathbf{1}_\text{IV}})}{\sqrt{2}}.
\label{BSeq1}
\end{equation}
If the photon is transmitted (reflected), it impinges on the cavity $A$ ($B$), performing a CPF gate ($\ket{\pm}\ket{\mathbf{1}}_\text{in} \to \ket{\mp}\ket{\mathbf{1}}_\text{out}$, $\ket{\pm}\ket{\mathbf{0}}_\text{in} \to \ket{\pm}\ket{\mathbf{0}}_\text{out}$ for an ideal gate), which yields
\begin{align}
&\ket{\!++}_{AB}\frac{(\ket{\mathbf{1}_\text{III},\mathbf{0}_\text{IV}}+i\ket{\mathbf{0}_\text{III},\mathbf{1}_\text{IV}})_\text{in}}{\sqrt{2}} \to \nonumber \\
&\frac{1}{\sqrt{2}}(\ket{\!-+}_{AB}\ket{\mathbf{1}_\text{III},\mathbf{0}_\text{IV}}_\text{out} + i\ket{\!+-}_{AB}\ket{\mathbf{0}_\text{III},\mathbf{1}_\text{IV}}_\text{out}),
\label{CPFremote}
\end{align}
with both cavities as well as the channels I and II of the BS remaining in the vacuum state. Finally, the photon pass through the BS again \cite{Campos1989}, 
\begin{align}
&\text{BS} \ket{\mathbf{0}_\text{I},\mathbf{0}_\text{II}}\ket{\mathbf{1}_\text{III},\mathbf{0}_\text{IV}} = \frac{(\ket{\mathbf{1}_\text{I},\mathbf{0}_\text{II}}-i\ket{\mathbf{0}_\text{I},\mathbf{1}_\text{II}})}{\sqrt{2}}\ket{\mathbf{0}_\text{III},\mathbf{0}_\text{IV}}, \\
&\text{BS} \ket{\mathbf{0}_\text{I},\mathbf{0}_\text{II}}\ket{\mathbf{0}_\text{III},\mathbf{1}_\text{IV}} = \frac{(-i\ket{\mathbf{1}_\text{I},\mathbf{0}_\text{II}}+\ket{\mathbf{0}_\text{I},\mathbf{1}_\text{II}})}{\sqrt{2}}\ket{\mathbf{0}_\text{III},\mathbf{0}_\text{IV}},
\label{BSeq2}
\end{align}
and then the photon is detected by one of the detectors ($D_1$ or $D_2$) [see Fig.~\ref{fig:5}]. If a photon count is registered in $D_1$, the atoms are projected into the state
\begin{equation}
\ket{\Psi_{D_1}} = \frac{1}{\sqrt{2}}(\ket{22}-\ket{11}),
\label{psiD1}
\end{equation}
while for a click in $D_2$ we have
\begin{equation}
\ket{\Psi_{D_2}} = \frac{1}{\sqrt{2}}(\ket{21}-\ket{12}).
\label{psiD2}
\end{equation}
Therefore, the atoms become maximally entangled whenever a photon count is registered in either $D_1$ or $D_2$. 

It is worth noticing that in our propose the single photon neither impinges on the cavities in sequence (serial quantum circuit) nor undergoes single-qubit operations as extensively adopted in literature. Here, on the other hand, we exploit the indistinguishability of the photon quantum paths, such that the interference between them can virtually work as the situation in which the single photon impinges on both cavities simultaneously. Besides of being a very simple single-step process to entangle remote atoms, our parallel quantum circuit also avoids the accumulation of errors due to the multiple (CPF and single-qubit) gates required in the serial approach. The above results [Eqs.~\eqref{psiD1}$-$\eqref{psiD2}] was achieved by considering an ideal CPF gate, but, as discussed in the first protocol, the efficiency and fidelity of a CAPS-based CPF gate depend on the cooperativity and the pulse duration, case which we will analyze in the following.

Considering the same assumptions made for the first protocol, the effective dynamics after the single photon crossing the BS and before passing through it again is given by $H_\text{eff}=H_\text{eff}^\text{III(A)} \oplus H_\text{eff}^\text{IV(B)}$. In an interaction picture rotating at the cavities resonance frequency, we have
\begin{align} \label{H2nd}
H_\text{eff}^{m(j)} &= \int_{-\infty}^{\infty} \!\!\!\! d\omega \, \omega b^{\dagger}_m(\omega)b_m(\omega)  \nonumber \\
&+ i\frac{\sqrt{2\kappa}}{\sqrt{2\pi}}\int_{-\infty}^{\infty} \!\!\!\! d\omega \,[a^{\dagger}_j b_m(\omega) - a_j b^{\dagger}_m(\omega)] \nonumber \\
&+g(a_j\sigma_{31}^{j} + a^{\dagger}_j\sigma_{13}^{j}) - i\Gamma_{3}\sigma_{33}^{j},
\end{align} 
where the superscripts $m$ and $j$ refers to the channel of the BS and atom-cavity system, respectively. 
As the atom-cavity systems evolve independently and are considered identical, we have to solve only the problem of a single-photon pulse impinging on a cavity with a single atom trapped in it, which is essentially the same that was done in the first protocol, except for neglecting one of the atoms in Eq.~\eqref{eq:1}. If the external multimode field is in the vacuum state with the atom-cavity system deexcited, there is no dynamics, i.e., the entire system remains in its global ground state. If the atom-cavity system has only one excitation and it is initially in the external field, the general evolved state for this system is
\begin{align} \label{psi_t2nd}
\ket{\psi(t)}_{m(j)} &=  c_{3,0}(t) \ket{3}_{j} \ket{0}_{c_j}  \ket{\mathbf{0}_m}  \nonumber \\
 & +\sum_{k=1}^{2} c_{k,1}(t)\ket{k}_{j} \ket{1}_{c_j}  \ket{\mathbf{0}_m}  \nonumber \\
 & + \sum_{k=1}^{2}\int_{-\infty}^{\infty} \!\!\!\! d\omega \, \xi_{k}(\omega,t)\ket{k}_{j} \ket{0}_{c_j}  b_m^{\dagger}(\omega)\ket{\mathbf{0}_m},
\end{align}
with the dynamics of $\{c_{1,1},c_{3,0},\xi_{1}\}$ and $\{c_{2,1},\xi_{2}\}$ being exactly given by Eq.~\eqref{eq:4c} and Eq.~\eqref{eq:4d}.

When the outgoing pulse is already far enough from the cavity ($t \gg 1$) and right before crossing the BS again
\begin{equation} \label{psi_t2ndss}
\ket{\psi_\text{ss}}_{m(j)} = \sum_{k=1}^{2}\int_{-\infty}^{\infty} \!\!\!\! d\omega \, \xi_\text{out}^{k}(\omega)\ket{k}_{j} \ket{0}_{c_j} \ket{\mathbf{1}_m},
\end{equation}
with $\xi_\text{out}^{k}(\omega) = \xi_{k}(\omega,t \gg 1)$ and $\ket{\mathbf{1}_m}=b_m^{\dagger}(\omega)\ket{\mathbf{0}_m}$. Therefore, assuming a perfect BS, the generalization of the ideal case discussed before is accomplished just by replacing in the right-hand side of Eq.~\eqref{CPFremote} 
\begin{equation} \label{repgen}
\frac{1}{\sqrt{2}}\ket{-}_{j} \ket{\mathbf{1}_m}_\text{out} \to \sum_{k=1}^{2}\int_{-\infty}^{\infty} \!\!\!\! d\omega \, \xi_\text{out}^{k}(\omega)\ket{k}_{j}  \ket{\mathbf{1}_m}_\text{out}.
\end{equation}
Thus, after the single-photon pulse crossing the BS again, the normalized atomic steady states with a photon count being registered in $D_1$ and $D_2$ are, respectively, 
\begin{align}
&\rho_{D1} = \frac{1}{P_{D_1}}\sum_{k,\ell,p,q=1}^{2}   \int_{-\infty}^{\infty} \!\!\!\! dt \, \beta_{\text{out}}^{k\ell}(t) \, \beta_{\text{out}}^{pq}(t)^{*} \ket{k\ell}\bra{pq},\label{rho1} \\
&\ket{\Psi_{D_2}} =\frac{1}{\sqrt{2}}(\ket{21}-\ket{12}), \label{psi2}
\end{align}
with $\beta_{\text{out}}^{k\ell}(t) = [\alpha_{\text{out}}^{k}(t)+\alpha_{\text{out}}^{\ell}(t)]/2$, remembering that $\alpha_{\text{out}}^{k}(t)$ is the Fourier transform of $\xi_\text{out}^{k}(\omega)$ and is determined by the input-output relation $\alpha_{\text{out}}^{k}(t) = \sqrt{2\kappa} \, c_{k,1}(t) - \alpha_{\text{in}}^{k}(t)$. The probabilities of registering a photon count in $D_1$ and $D_2$ are, respectively,
\begin{align}
P_{D_1} &= \sum_{k,\ell=1}^{2} \int_{-\infty}^{\infty} \!\!\!\! dt \, |\beta_{\text{out}}^{k\ell}(t)|^{2},\label{Ps1} \\
P_{D_2} &= \frac{1}{2} \int_{-\infty}^{\infty} \!\!\!\! dt \, |\alpha_{\text{out}}^{2}(t) - \alpha_{\text{out}}^{1}(t)|^{2}. \label{Ps2}
\end{align}

Equation \eqref{psi2} provides a remarkable result. Namely, the atoms become always maximally entangled as long as a photon is detected in $D_2$. The more interesting point is that this happens regardless the value of the cooperativity and the pulse duration. Therefore, the mechanism behind this specific entanglement generation is not the CPF gate, but the symmetry of the problem together with the interference of the quantum optical paths via the BS. However, although the entanglement degree does not depend on the cooperativity in this case, the probability of registering a photon count in $D_2$ does. For the cases in which the photon is detected in $D_1$, we have a similar result to that obtained in the first protocol [compare Eqs.~\eqref{rhoatss} and \eqref{rho1}].

\begin{figure}[t]
\includegraphics[trim = 14mm 28mm 18mm 35mm, clip, width=0.48\textwidth]{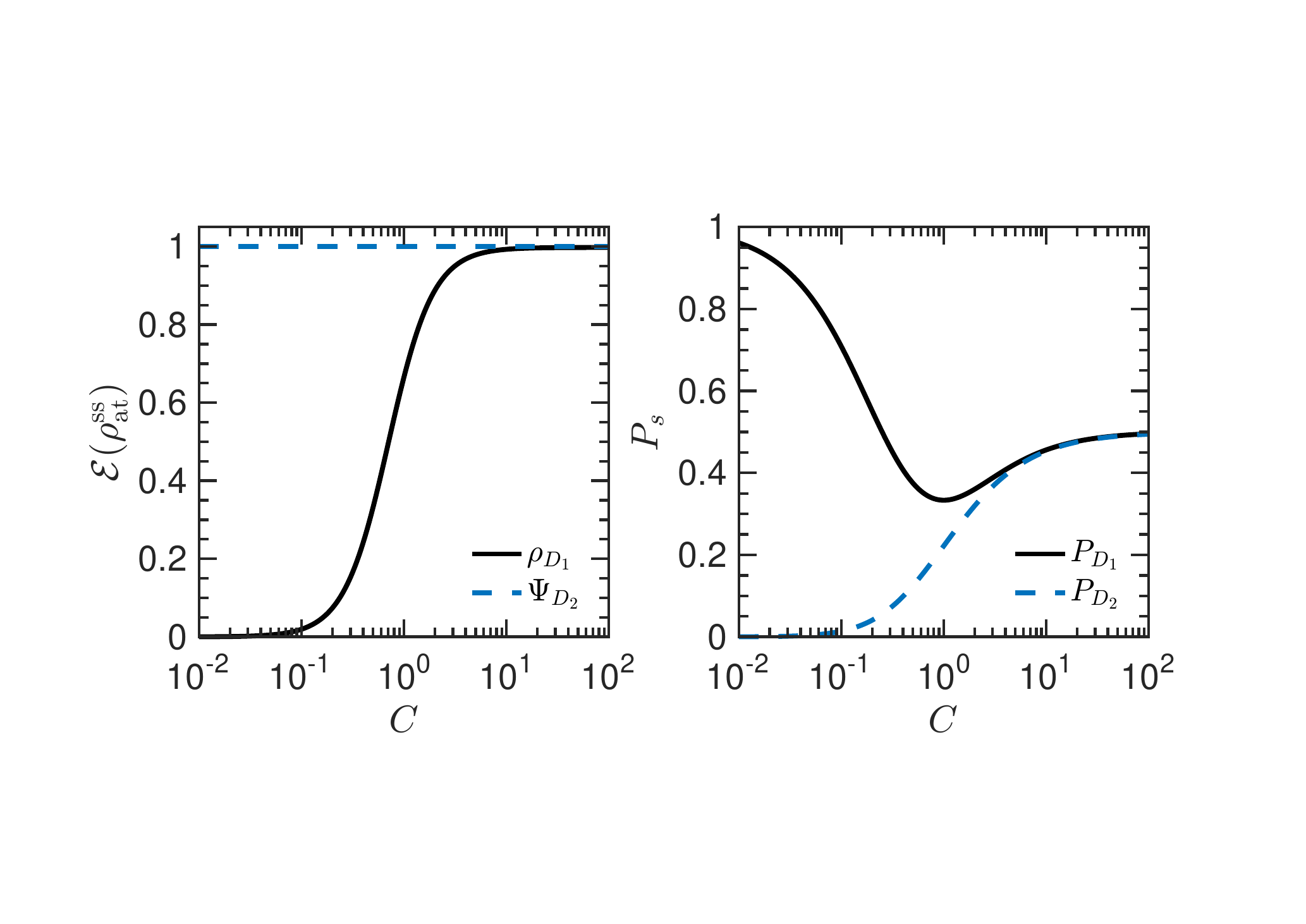}
\caption{(a) Concurrence associated to the atomic steady states $\rho_{D1}$ (solid black line) and $\ket{\Psi_{D_2}}$ (blue dashed line), respectively, as a function of the cooperativity. (b) Detection probability of registering a photon count in $D_1$ (solid black line) and $D_2$ (blue dashed line), respectively, as a function of the cooperativity. He we consider a long pulse duration, $\tau_{p}=50\kappa^{-1}$ and the initial state given by the left-hand side of Eq.~\eqref{CPFremote} [$\alpha_{\text{in}}^{k}(t)=\alpha_{\text{in}}(t)/2$].}
\label{fig:6}
\end{figure}

Let us assume $\Gamma_3 = \kappa$ again, but without loss of generality, in order to obtain the compact semianalytical solution
\begin{align}\label{aout_analy_rem}
{\alpha_{\text{out}}^{k}(t)} = 2\kappa \! &\int_{-\infty}^{t} \!\!\!\! ds \,  \cos{[g\delta_{k,1}(t-s)]} \nonumber \\
&\times \alpha_{\text{in}}^{k}(s)e^{-\kappa(t-s)} - \alpha_{\text{in}}^{k}(t).
\end{align}
Furthermore, we also consider the same input pulse of Eq.~\eqref{shape} in order to compute the entanglement degree and the probability for achieving it associated to each possible atomic steady state, which are shown in Fig.~\ref{fig:6} considering a large pulse duration ($\tau_{p} = 50\kappa^{-1}$) and the initial state given by the left-hand side of Eq.~\eqref{CPFremote} ($\alpha_{\text{in}}^{k}(t)=\alpha_{\text{in}}(t)/2$). We observe that the atoms are maximally entanglement for any positive value of cooperativity whenever the photon is detected in $D_2$. Nonetheless, the probability of registering a photon count in this detector is very low for small values of cooperativity and asymptotically saturates to $50\%$ for high cooperativities. On the other hand, if photon is detected in $D_1$, the concurrence is maximum only asymptotically in $C$, with the detection probability also saturating to $50\%$ as $C$ increases. 

It is worth emphasizing that our protocol predicts very high fidelity and efficiency in entangling two remote atoms when we consider the state-of-the-art parameters of very similar experimental setups ($C \sim 3$) \cite{Reiserer2014,Kalb2015,Hacker2016}. Namely, for this value of cooperativity, we theoretically obtain $\mathcal{E}(\rho_{D1}) \simeq 0.95$ with $P_{D_1} \simeq 39\%$ and $\mathcal{E}(\ket{\Psi_{D_2}}) = 1$ with $P_{D_2} \simeq 37\%$, which provides a total efficiency of $P_\text{total} \simeq 76\%$ for obtaining at least $95\%$ of entanglement (considering ideal photodetectors). Moreover, besides being a parallel quantum circuit instead of the commonly-adopted serial one, our protocol has also the advantage of not requiring both the interference and the simultaneous detection of two photons emitted from the two respective atoms \cite{Barret2005,Moer2007,Hoff2012,Bernien2013,Casabone2013,Delteil2016}, neither that one atom absorbs a single photon emitted by the other atom \cite{Ritter2012}.

Finally, our very simple single-step process presented here can also be straightforwardly applied to entangle two distant macroscopic atomic clouds. Consider that in Fig.~\ref{fig:5} we have a cloud with $N_A$ ($N_B$) atoms in cavity $A$ ($B$). In order to be succinct, let us assume that the CPF gates are almost perfectly performed, since the CAPS-based CPF gate has higher fidelity and efficiency as the number of atoms inside the cavity increases \cite{Duan2005}. For instance, if we initially prepare each atomic cloud in a GHZ state, $\ket{\theta(\varphi)}_j=(\ket{1}^{\otimes N_j} + e^{i\varphi}\ket{2}^{\otimes N_j})/\sqrt{2}$ \cite{Chen2015,Reiter2016,Shao2017}, it is easy to show that the atomic state, by sending a single photon to the BS and then detecting it after passing through the BS again, is given by
\begin{equation} \label{cloud}
\ket{\Phi_{\pm}} = \frac{\ket{\theta(\varphi)}_A \ket{\theta(\phi)}_B \pm \ket{\theta(\varphi-\pi)}_A \ket{\theta(\phi-\pi)}_{B}}{\sqrt{2}},
\end{equation} 
with the plus (minus) sign standing for the case in which the photon is registered in $D_1$ ($D_2$). For this case we see that the atomic clouds (macroscopic objects) become maximally entangled with each other. Although we can find protocols for generating GHZ states elsewhere \cite{Chen2015,Reiter2016,Shao2017}, they often become a laborious task as the number of atoms increases. On the other hand, our results pave the way to interesting future studies for entangling macroscopic clouds of atoms, e.g., by using more accessible initial states or by adding further steps to the process.

\section{Conclusion}
\label{sec:4}
We have investigated two protocols of heralded-entanglement generation between two atoms based on cavity-assisted photon scattering. Here it was performed a detail study on which conditions the degree of entanglement can be optimized considering either both atoms inside the same cavity or each one trapped in distant cavities. The key ingredient of our proposal is a controlled-phase-flip gate where a phase shift can be imprinted on the output field depending on the atomic state. For both protocols, our results showed that the entanglement degree and the success probability are close to unity in the strong-coupling regime reached in the current technologies. The great advantage of our proposals is the entanglement generation through a very simple single-step process which minimizes the sources of error, increasing the efficiency with less resources. For atoms trapped in distant cavities, we introduce a \textit{quantum parallel circuit} instead of the serial process extensively adopted in the literature. This very simple parallel circuit can straightforwardly applied to entangle two distant macroscopic atomic clouds. Among other applications, our proposal and its extension to multiple atom-cavity systems step toward a suitable route for quantum networking, in particular for quantum state transfer, quantum teleportation and nonlocal quantum memory.

\section{Acknowledgements}

This work was supported by the S\~{a}o Paulo Research Foundation (FAPESP) Grants No.~2013/04162-5, No.~2013/23512-7, and No.~2014/12740-1, National Council for Scientific and Technological Development (CNPq) Grant No.~150879/2017-2, and Brazilian National Institute of Science and Technology for Quantum Information (INCT-IQ).

\appendix

\section{Steady state with two atoms inside the cavity}
\label{appA}

Considering two atoms inside the same cavity (Sec.~\ref{sec:two}), with the general evolved state and the effective Hamiltonian described by Eqs.~\eqref{psi_t} and \eqref{heff}, respectively, the dynamics of the system is given by the Schr\"{o}dinger equation, $ i \partial_{t}\ket{\psi(t)} = H_{\text{eff}}\ket{\psi(t)}$ ($\hbar =1$), which yields the following sets of coupled integro-differential equations for the amplitude coefficients
\begin{subequations}
\label{eq:4}
\begin{align}
\begin{pmatrix} \label{eq:4a}
\dot{c}_{11,1} \\ 
\dot{c}_{13,0}\\ 
\dot{c}_{31,0}\\ 
\dot{\xi}_{11}\\
\end{pmatrix} = 
\begin{pmatrix}
0 & -ig & -ig & \sqrt{\kappa/\pi}\int \!\! d\omega\\ 
-i g & -\Gamma_3  & 0 & 0 \\ 
-i g & 0 & -\Gamma_3 &  0 \\ 
-\sqrt{\kappa/\pi} & 0 & 0 &  -i \omega \\ 
\end{pmatrix}
\begin{pmatrix} 
c_{11,1} \\ 
c_{13,0}\\ 
c_{31,0}\\ 
\xi_{11}\\
\end{pmatrix},\\
\begin{pmatrix}\label{eq:4b}
\dot{c}_{12,1} \\ 
\dot{c}_{32,0}\\  
\dot{\xi}_{12}\\
\end{pmatrix} = 
\begin{pmatrix}
0& -ig & \sqrt{\kappa/\pi}\int \!\! d\omega\\ 
-i g & -\Gamma_3  & 0  \\ 
-\sqrt{\kappa/\pi} & 0 &  -i \omega \\ 
\end{pmatrix}
\begin{pmatrix} 
c_{12,1} \\ 
c_{32,0}\\  
\xi_{12}\\
\end{pmatrix},\\
\begin{pmatrix} \label{eq:4c}
\dot{c}_{21,1} \\ 
\dot{c}_{23,0}\\  
\dot{\xi}_{21}\\
\end{pmatrix} = 
\begin{pmatrix}
0& -ig & \sqrt{\kappa/\pi}\int \!\! d\omega\\ 
-i g & -\Gamma_3  & 0  \\ 
-\sqrt{\kappa/\pi} & 0 &  -i \omega \\ 
\end{pmatrix}
\begin{pmatrix}
c_{21,1} \\ 
c_{23,0}\\  
\xi_{21}\\
\end{pmatrix},\\
\begin{pmatrix} \label{eq:4d}
\dot{c}_{22,1} \\   
\dot{\xi}_{22}\\
\end{pmatrix} = 
\begin{pmatrix}
0& \sqrt{\kappa/\pi}\int \!\! d\omega\\ 
-\sqrt{\kappa/\pi} &  -i \omega \\ 
\end{pmatrix}
\begin{pmatrix}
c_{22,1} \\ 
\xi_{22}\\
\end{pmatrix}
.
\end{align}
\end{subequations}

Integrating $\dot{\xi}_{k\ell}=-\sqrt{(\kappa/\pi)} c_{k\ell,1}  -i \omega \xi_{k\ell}$ from some past time $t_0$, when the incoming single-photon pulse is still at a sufficiently large distance from the cavity, to a time $t>t_0$, we obtain
\begin{align} \label{xi_in}
\xi_{k\ell}(\omega,t) = &\overbrace{\xi_{k\ell}(\omega,t_0)}^{{\xi_{\text{in}}^{k\ell}}\!(\omega)}e^{-i\omega(t-t_0)} \nonumber \\
 &- \sqrt{\frac{\kappa}{\pi}}\int_{t_0}^{t} \!\!\! d\tau \, c_{k\ell,1}(\tau)e^{-i\omega(t-\tau)}.
\end{align}
%
Inserting the above result into $\dot{c}_{k\ell,1} \propto \sqrt{\kappa/\pi}\int \!\! d\omega \, \xi_{k\ell}(\omega,t)$, we have
\begin{align} \label{ckl}
\sqrt{\frac{\kappa}{\pi}}\int \!\! d\omega \, \xi_{k\ell}(\omega,t)&= \sqrt{\frac{\kappa}{\pi}} \overbrace{\int_{-\infty}^{\infty} \!\!\! d\omega \, \xi_{\text{in}}^{k\ell}(\omega)e^{-i\omega(t-t_0)}}^{\sqrt{2\pi}\alpha_{\text{in}}^{k\ell}(t)} \nonumber \\
 & -\frac{\kappa}{\pi}  \underbrace{\int_{t_0}^{t} \!\!\! d\tau \, c_{k\ell,1}(\tau) \!\! \underbrace{\int_{-\infty}^{\infty} \!\!\! d\omega e^{-i\omega(t-\tau)}}_{2\pi\delta(t-\tau)}}_{\pi c_{k\ell,1}(t) } \nonumber\\
&= -\kappa \, c_{k\ell,1}(t)+ \sqrt{2\kappa}\,\alpha_{\text{in}}^{k\ell}(t).
\end{align}
With this, the integro-differential equations in Eq.~\eqref{eq:4} are reduced to ordinary differential equations, which are equivalent to those that will be obtained by employing the method of Ref.~\cite{Kuhn12}.

On the other hand, instead of integrating $\dot{\xi}_{k\ell}$ from a past time $t_0$ to $t$, we can also integrate it from $t$ to a future time $t_1>t$, when the outgoing pulse is already far enough from the cavity. For this case, 
\begin{align} \label{xi_out}
\xi_{k\ell}(\omega,t) = &\overbrace{\xi_{k\ell}(\omega,t_1)}^{-{\xi_{\text{out}}^{k\ell}}\!(\omega)}e^{-i\omega(t-t_1)} \nonumber \\
 &+ \sqrt{\frac{\kappa}{\pi}}\int_{t}^{t_1} \!\!\! d\tau \, c_{k\ell,1}(\tau)e^{-i\omega(t-\tau)},
\end{align}
where the minus sign in $\xi_{k\ell}(\omega,t_1) = -{\xi_{\text{out}}^{k\ell}}(\omega)$ comes from the convention that explicitly incorporates the propagation direction of the incoming and outgoing fields in their amplitudes \cite{Walls2007}. By taking the Fourier transform of the combination of Eqs.~\eqref{xi_in} and \eqref{xi_out}, we obtain the input-output relation
\begin{equation} \label{io_rel}
\alpha_{\text{out}}^{k\ell}(t) = \sqrt{2\kappa} \, c_{k\ell,1}(t) - \alpha_{\text{in}}^{k\ell}(t),
\end{equation}
which represents a boundary condition relating the far field amplitudes outside the cavity to the intracavity field. 

From these results, we are able to calculate the atomic steady state after detecting the outgoing photon. Considering the Eq.~\eqref{psi_t} and given $\tilde{\rho}(t) = \ket{\psi(t)} \bra{\psi(t)}$, the reduced atomic density matrix when there is a photon in the output field (reservoir) is
\begin{align} \label{rhoat_unn}
\tilde{\rho}_{\text{at}}(t) &= \text{Tr}_{\text{cav}+\text{res}} \left[ \textstyle\int_{-\infty}^{\infty} \! d\nu \,b^{\dagger} (\nu)\ket{\mathbf{0}} \bra{\mathbf{0}} b(\nu)  \tilde{\rho}(t) \right] \nonumber \\
&=\sum_{m=0}^{1} \int_{-\infty}^{\infty} \!\!\!\! d\nu \, \bra{m}_{c}  \bra{\mathbf{0}} b(\nu) \, \tilde{\rho}(t) \, b^{\dagger} (\nu)\ket{\mathbf{0}}  \ket{m}_{c} \nonumber \\
& = \sum_{k,\ell=1}^{2} \sum_{p,q=1}^{2} \int_{-\infty}^{\infty} \!\!\!\! d\omega \, \xi_{k\ell}(\omega,t) \xi_{pq}^{*}(\omega,t) \ket{k\ell}\bra{pq},
\end{align}
in which the tilde means that the density matrix is unnormalized.

Since the photon detection must occur when the outgoing field is far enough from the cavity ($t \to t_1$), i.e., when the pulse and the atoms-cavity system no longer interact, the normalized atomic steady state, $\rho_{\text{at}}^{\text{ss}} = \lim_{t \to t_1}\{\tilde{\rho}(t)/\text{Tr}[\tilde{\rho}(t)]\}$, is given by
\begin{equation} \label{apprho}
\rho_{\text{at}}^{\text{ss}} = \frac{1}{P_s}\sum_{k,\ell=1}^{2} \sum_{p,q=1}^{2} \int_{-\infty}^{\infty} \!\!\!\! dt \, \alpha_{\text{out}}^{k\ell}(t) \, \alpha_{\text{out}}^{pq}(t)^{*} \ket{k\ell}\bra{pq},
\end{equation}
where we use $\xi_{k\ell}(\omega,t_1) = -{\xi_{\text{out}}^{k\ell}}(\omega)$ and the fact that ${\xi_{\text{out}}^{k\ell}}(\omega)$ is the Fourier transform of ${\alpha_{\text{out}}^{k\ell}}(t)$. The normalization factor in the above equation, 
\begin{equation}
P_s \equiv \sum_{k,\ell=1}^{2} \int_{-\infty}^{\infty} \!\!\!\! dt \, |\alpha_{\text{out}}^{k\ell}(t)|^{2},
\end{equation}
is exactly the average number of photons in the output field after a long time, i.e., it gives the probability of the (ideal) detector in Fig.~\ref{fig:1} register a photon count, which sets the success probability of the entangling gate.


\begin{thebibliography}{99}
\bibitem{rotcarr} In the original frame, $\tilde{\xi}(\omega)$ is written as a slowly varying envelope rotating at the carrier frequency, $\tilde{\xi}(\omega)=\xi(\omega)e^{-i\omega_c t}$, in which $\xi(\omega)$ is centered on the carrier frequency $\omega_c$. On the other hand, in the interaction picture rotating at $\omega_c$, $\xi(\omega)$ becomes a function centered on $\omega = 0$ (equivalent to $\omega = \omega_c$ in the original frame) due to a variable subtitution in the integrals $(\omega \rightarrow \omega + \omega_c)$.
\end{thebibliography}
\end{document}